\documentclass[reqno]{amsart}

\newcommand{\Rm}{\mathbb{R}}
\newcommand{\Cm}{\mathbb{C}}

\newcommand{\Sm}{\mathbb{S}}
\newcommand{\be}{\begin{equation}}
\newcommand{\ee}{\end{equation}}
\newcommand{\ba}{\begin{equation}\begin{aligned}}
\newcommand{\ea}{\end{aligned}\end{equation}}
\newcommand{\va}{\varphi}
\newcommand{\vth}{\vartheta}
\newcommand{\pp}{\partial}
\newcommand{\vv}[1]{\boldsymbol{\mathrm{#1}}}
\newcommand{\vs}[1]{\boldsymbol{#1}}
\newcommand{\hvv}[1]{\boldsymbol{\hat{\mathrm{#1}}}}
\newcommand{\uv}{\boldsymbol{{\hat{\mathrm{s}}}}}
\newcommand{\uvk}{\boldsymbol{\hat{\mathrm{k}}}}
\newcommand{\ket}[1]{\left|#1\right\rangle}
\newcommand{\bra}[1]{\left\langle#1\right|}
\newcommand{\braket}[2]{\left\langle#1\middle|#2\right\rangle}

\newcommand{\rrf}[1]{\mathop{\mathcal{R}_{{#1}}}}

\newcommand{\argmin}{\mathop{\mathrm{arg\,min}}}

\usepackage{amsmath}
\usepackage[foot]{amsaddr}
\usepackage{amsthm,amssymb,mathrsfs}
\usepackage{graphicx}
\usepackage{color}

\theoremstyle{remark}

\title[]{Decay behavior and optical parameter identification for spatial-frequency domain imaging by the radiative transport equation}

\author{Manabu Machida$^1$}
\author{Yoko Hoshi$^1$}
\author{Keiichiro Kagawa$^2$}
\author{Kazuki Takada$^3$}

\address{$^1$ Institute for Medical Photonics Research, Hamamatsu University School of Medicine, Hamamatsu 431-3192, Japan}
\address{$^2$ Research Institute of Electronics, Shizuoka University, Hamamatsu 432-8011, Japan}
\address{$^3$ Graduate School of Integrated Science and Technology, Shizuoka University, Hamamatsu 432-8011, Japan}
\email{machida@hama-med.ac.jp (M. Machida)}

\begin{document}

\begin{abstract}
The decay behavior of the specific intensity is studied for the spatial-frequency domain imaging (SFDI). It is shown using the radiative transport equation that the decay is given by a superposition of different decay modes, and the decay rates of these modes are determined by spatial frequencies and Case's eigenvalues. This explains why SFDI can focus on shallow regions. The fact that light with nonzero spatial frequency rapidly decays makes it possible to exclusively extract optical properties of the top layer of a layered medium. We determine optical properties of the top layer of a solid phantom. This measurement is verified with different layered media of numerical phantoms.
\end{abstract}

\maketitle

\section{Introduction}
\label{intro}

In near-infrared spectroscopy, light illumination in spatial-frequency domain has been developed as a tool which is concise as the continuous-wave illumination and informative as the frequency-domain illumination \cite{Gioux-Mazhar-Cuccia19}. The spatial-frequency domain imaging (SFDI) is capable of determining both absorption and scattering coefficients from time-independent measurements. SFDI is mainly used to extract optical properties at depths of the order of millimeters. SFDI was used for imaging skin flap oxygenation during reconstructive breast surgery \cite{Gioux-etal11}. It was also used to record biochemical compositional changes in port wine stain after laser therapy \cite{Mazhar-etal12}. Tissue optical properties of a human volar forearm were estimated by SFDI \cite{Nadeau-etal14}. Burn wounds were examined by SFDI coupled with laser speckle imaging \cite{Ponticorvo-etal14}. See a recent review by Angelo, et al.~\cite{Angelo-etal18} and references therein.

The above mentioned works show that SFDI can exclusively study shallow regions near the skin. In SFDI, spatially modulated incident beams rapidly decay in biological tissue \cite{Cuccia-etal09}. In this paper, we further investigate this feature of SFDI. This property of SFDI is advantageous when we are interested in measuring optical properties of shallow regions. Even when the thickness is thin, we can assume the half space, which is unbounded in the depth direction. In this paper, we identify optical properties of the top layer of a layered medium. It is not possible to extract optical properties of the top layer in the standard setting of near-infrared spectroscopy, in which optical fibers are attached on the top of a layered medium, because near-infrared light propagates not only in the top layer but reaches deeper layers.

In this paper, the ability to extract optical properties of the top layer is tested by different numerical phantoms. Moreover, optical properties of the top layer of a solid phantom is determined. As a numerical tool for this parameter identification, we demonstrate that the numerical scheme for the radiative transport equation (RTE) based on the method of rotated reference frames \cite{Markel04,Panasyuk-etal06} provides an efficient numerical algorithm for SFDI partially because the method relies on the Fourier transform in the spatial frequency domain.

Noting the fact that the solution to the radiative transport equation are expressed as a superposition of three-dimensional singular eigenfunctions \cite{Machida14}, we investigate the asymptotic behavior of the solution. The effect of the spatial frequency $q_0$ on the decay of the specific intensity is found. The longest lived mode is controlled by $q_0$ and the largest Case's eigenvalue. This finding gives a theoretical reason why shallow regions can be exclusively studied by SFDI.

Then we numerically solve the radiative transport equation in the half space by the method of rotated reference frames \cite{Markel04}. The method of rotated reference frames in the half space was developed for a point source \cite{Machida-etal10,Panasyuk-etal06} and for a spatially oscillating source \cite{Liemert-Kienle12}. In \cite{Panasyuk-etal06,Liemert-Kienle13}, the subtraction of the ballistic term was considered. Using the method of rotated reference frames, the effect of surface scattering in SFDI was studied \cite{Nothelfer-etal19}. The inverse problem is solved by the Levenberg-Marquardt algorithm.  With our approach, optical parameters of the top layer of a layered solid phantom made of epoxy resin was determined within $1$ sec on a laptop computer.

The remainder of this paper is organized as follows. In Sec.~\ref{asym}, we study the asymptotic behavior of the specific intensity to consider the penetration depth of near-infrared light illuminated by a spatially modulated source. In Sec.~\ref{sfdi}, reconstruction of optical properties of the top layer of layered media is considered for different numerical phantoms. Optical properties are estimated using a solid phantom in Sec.~\ref{experiments}. Discussion and conclusions are given in Sec.~\ref{concl}. In Appendix \ref{rte}, two types of the diffusion approximation are introduced. Appendix \ref{mrrfmain} is devoted to the eigenmode expansion and numerical algorithm of the RTE.

\section{The decay behavior}
\label{asym}

Let us consider near-infrared light propagation in the half space. Let $\Omega$ be the half space , i.e.,
\be
\Omega=\left\{\vv{r}\in\Rm^3;\;-\infty<x<\infty,\;-\infty<y<\infty,\;0<z<\infty\right\}.
\ee
Let $\pp\Omega$ be the boundary of $\Omega$, i.e., the $x$-$y$ plane. The specific intensity at position $\vv{r}$ ($\vv{r}=(\vs{\rho},z)^T$, $\vs{\rho}=(x,y)^T$) in direction $\uv$ is denoted by $I(\vv{r},\uv)$, where $\uv$ is a unit vector specified by the polar angle $\vth\in[0,\pi]$ and azimuthal angle $\va\in[0,2\pi)$. Let $d\uv$ denote $\sin\vth\,d\vth d\va$. The RTE is written as
\be
\left\{\begin{aligned}
\left(\uv\cdot\nabla+\mu_t\right)I(\vv{r},\uv)=
\mu_s\int_{\Sm^2}p(\uv,\uv')I(\vv{r},\uv')\,d\uv',
\\
(\vv{r},\uv)\in\Omega\times\Sm^2,
\\
I(\vv{r},\uv)=
R_{\mathfrak{n}}(\uv\cdot\hvv{z})I(\vv{r},\uv_R)+I_{\rm inc}(\vv{r},\uv),
\quad (\vv{r},\uv)\in\Gamma_-,
\end{aligned}\right.
\label{rte0}
\ee
where total attenuation $\mu_t$ is the sum of absorption coefficient $\mu_a$ and scattering coefficient $\mu_s$, which are both assumed to be positive constants, and $p(\uv,\uv')$ is the scattering phase function. We assume that $p(\uv,\uv')$ is given by
\be
p(\uv,\uv')=
\sum_{l=0}^{l_{\rm max}}\sum_{m=-l}^l{\rm g}^lY_{lm}(\uv)Y_{lm}^*(\uv'),
\label{phasefunc}
\ee
where $l_{\rm max}$ is a positive integer, ${\rm g}\in(-1,1)$ is a constant, and the superscript $*$ denotes complex conjugate. Spherical harmonics $Y_{lm}(\uv)$ are defined by
\be
Y_{lm}(\uv)=\sqrt{\frac{2l+1}{4\pi}\frac{(l-m)!}{(l+m)!}}P_l^m(\cos\vth)e^{im\va},
\ee
where $P_l^m(\mu)$ are associated Legendre polynomials. Throughout the paper, we set $l_{\rm max}=9$. Moreover,
\be
\begin{aligned}
\Gamma_-
&=
\left\{(\vv{r},\uv)\in\pp\Omega\times\Sm^2;\;\vs{\nu}(\vv{r})\cdot\uv<0\right\}
\\
&=
\left\{(\vv{r},\uv)\in\pp\Omega\times\Sm^2_+\right\},
\end{aligned}
\ee
where $\vs{\nu}(\vv{r})$ is the outer unit vector normal to $\vv{r}\in\pp\Omega$ and $\Sm^2_+$ denotes the set of unit vectors in inward directions. We give the incident beam $I_{\rm inc}(\vv{r},\uv)$ as
\be
I_{\rm inc}(\vv{r},\uv)=e^{i\vv{q}_0\cdot\vs{\rho}}\delta(\uv-\hvv{z}),
\quad\vv{q}_0\in\Rm^2.
\ee
where $\hvv{z}$ be the unit vector in the positive $z$ direction. The Fresnel reflection for the ratio $\mathfrak{n}$ between the refractive indices inside and outside is also considered in the boundary condition. The direction $\uv_R$ is specified by the polar angle $\pi-\vth$ and azimuthal angle $\va$. Assuming unpolarized light, the Fresnel coefficient $R_{\mathfrak{n}}(\mu)$ ($0<\mu\le1$) is given by \cite{Aronson95}
\be
R_{\mathfrak{n}}(\mu)=\left\{\begin{aligned}
\frac{1}{2}\left(\left(\frac{\mu-\mathfrak{n}\mu_0}{\mu+\mathfrak{n}\mu_0}\right)^2+\left(\frac{\mu_0-\mathfrak{n}\mu}{\mu_0+\mathfrak{n}\mu}\right)^2\right)
&\quad\mbox{for}\;\mu\ge\mu_c,
\\
1
&\quad\mbox{for}\;\mu<\mu_c,
\end{aligned}\right.
\ee
where $\mu_0=\sqrt{1-\mathfrak{n}^2(1-\mu^2)}$ and $\mu_c=\sqrt{\mathfrak{n}^2-1}/\mathfrak{n}$.

Suppose that $\vv{r}_d$ is a point on $\pp\Omega$. We detect the hemispheric flux
\be
\begin{aligned}
J_+(\vv{r}_d)
&=
\int_0^{2\pi}\int_{\pi}^{2\pi}(\cos\vth)I(\vv{r}_d,\uv)\sin\vth\,d\vth d\va
\\
&=
-A_{\rm RTE}(q_0)e^{i\vv{q}_0\cdot\vs{\rho}},
\end{aligned}
\label{JplusA}
\ee
where $q_0=|\vv{q}_0|$ and $A_{\rm RTE}(q_0)$ is given by (\ref{Atheor2}) in Appendix \ref{mrrfmain}.

The fluence of the specific intensity is asymptotically governed by the diffusion equation. Let $u(\vv{r})$ denote the solution of the diffusion equation. To do the diffusion approximation, we split the specific intensity into two terms (see Appendix \ref{rte}): $I(\vv{r},\uv)=I_0(\vv{r},\uv)+I_1(\vv{r},\uv)$, where $I_0(\vv{r},\uv)$ satisfies
\be
\left\{\begin{aligned}
\left(\uv\cdot\nabla+\bar{\mu}\right)I_0(\vv{r},\uv)=0,
&\quad (\vv{r},\uv)\in\Omega\times\Sm^2,
\\
I_0(\vv{r},\uv)=I_{\rm inc}(\vv{r},\uv),
&\quad (\vv{r},\uv)\in\Gamma_-.
\end{aligned}\right.
\ee
Here, different choices are possible for $\bar{\mu}$ \cite{Machida-etal09,Tricoli-etal18}. Then $I_1$ is determined depending on the choice of $\bar{\mu}$. The $P_1$ approximation is made for this $I_1$ and eventually we arrive at the diffusion equation. We introduce
\be
\mu_s'=(1-{\rm g})\mu_s,\quad\mu_*=\mu_a+\mu_s',\quad
\mu_{\rm eff}=\sqrt{3\mu_a\mu_*}.
\ee
Probably the most naive choice is
\be
\bar{\mu}=\mu_a+\mu_s.
\ee
We call this diffusion approximation DA1. In this case, from (\ref{uandv}), (\ref{DA1B}), (\ref{vDA1BC}), and (\ref{DA1C}), we obtain
\be
u(\vv{r})=v_{\rm DA1}(z)e^{i\vv{q}_0\cdot\vs{\rho}},
\label{case1u}
\ee
where
\be
\begin{aligned}
v_{\rm DA1}(z)
&=\frac{3\mu_s(\mu_*+{\rm g}\mu_t)}{\mu_t^2-\mu_{\rm eff}^2-q_0^2}
\\
&
\times\left(\frac{\mu_t+3\mu_*/\zeta}{\sqrt{\mu_{\rm eff}^2+q_0^2}+3\mu_*/\zeta}e^{-\sqrt{\mu_{\rm eff}^2+q_0^2}z}-e^{-\mu_tz}\right).
\end{aligned}
\ee
Another choice is to set \cite{Svaasand-etal99}
\be
\bar{\mu}=\mu_*.
\ee
We refer to this diffusion approximation as DA2. In this case, from (\ref{uandv}), (\ref{vDA2BC}), and (\ref{DA2C}), we have
\be
u(\vv{r})=v_{\rm DA2}(z)e^{i\vv{q}_0\cdot\vs{\rho}},
\label{case2u}
\ee
where
\be
\begin{aligned}
v_{\rm DA2}(z)
&=
\frac{3\mu_s'\mu_*}{\mu_*^2-\mu_{\rm eff}^2-q_0^2}
\\
&\times
\left(\frac{\mu_*+3\mu_*/\zeta}{\sqrt{\mu_{\rm eff}^2+q_0^2}+3\mu_*/\zeta}e^{-\sqrt{\mu_{\rm eff}^2+q_0^2}z}-e^{-\mu_*z}\right).
\end{aligned}
\ee
Using the diffusion approximation, the detected light $J_+(\vv{r}_d)$ can be expressed as
\be
-A_{\rm DA1}(q_0)e^{i\vv{q}_0\cdot\vs{\rho}}\quad\mbox{or}\quad
-A_{\rm DA2}(q_0)e^{i\vv{q}_0\cdot\vs{\rho}},
\ee
depending on DA1 or DA2. They are given by (\ref{JpADA1}) or (\ref{JpADA2}) in Appendix \ref{rte}.

In the singular-eigenfunction approach \cite{Case-Zweifel}, the separation constant $\nu$ is either an eigenvalue $\nu_j(M)>1$ ($j=1,\dots,J^M$) or in the continuous spectrum $(0,1)$. Although there is only one positive eigenvalue $\nu_0$ in the case of isotropic scattering, in general there are multiple eigenvalues. We order them as $\nu_1>\nu_2>\cdots>\nu_{J^M}>1$ for each $M$. 

Case's method can be extended to three dimensions \cite{Machida14}, and the solution to (\ref{rte0}) can be expressed as
\be
\begin{aligned}
&
I(\vv{r},\uv)
=
e^{i\vv{q}_0\cdot\vs{\rho}}\sum_{M=-l_{\rm max}}^{l_{\rm max}}
\\
&\times
\Biggl[\sum_{j=0}^{J^M-1}
a_j^M\Psi_{\nu_j(M)}^M(\uv,\vv{q}_0)
e^{-\mu_t\hat{k}_z(\nu_j(M)q_0)z/\nu_j(M)}
\\
&+
\int_0^1
a^M(\nu)\Psi_{\nu}^M(\uv,\vv{q}_0)e^{-\mu_t\hat{k}_z(\nu q_0)z/\nu}\,d\nu
\Biggr]
\end{aligned}
\label{decaymodes}
\ee
with coefficients $a_j^M$, $a^M(\nu)$. Here, $\Psi_{\nu_j(M)}^M(\uv,\vv{q}_0),\Psi_{\nu}^M(\uv,\vv{q}_0)$ are three-dimensional singular eigenfunctions introduced in Appendix \ref{mrrfmain}.

We note that $\nu_0=\nu_1(0)>1$ is the largest eigenvalue. Let us define
\be
\begin{aligned}
I_0(\vv{r},\uv)
&=
I_0(\vs{\rho},z,\uv)
\\
&=
e^{i\vv{q}_0\cdot\vs{\rho}}a_0^0\Psi_{\nu_0}^0(\uv,\vv{q}_0)e^{-\mu_t\hat{k}_z(\nu_0q_0)z/\nu_0}.
\end{aligned}
\ee
When $z$ is large, the contribution of the mode $I_0$ dominates:
\be
\begin{aligned}
&
\left\|I(\cdot,z,\cdot)-I_0(\cdot,z,\cdot)\right\|_{L^{\infty}(\Rm^2;L^{\infty}(\Sm^2))}
\\
&=
o\left(\exp\left(-z\sqrt{\left(\frac{\mu_t}{\nu_0}\right)^2+q_0^2}\right)\right)
\end{aligned}
\label{mainthm}
\ee
as $z\to\infty$.

In the case of isotropic scattering ($g=0$) \cite{Case-Zweifel}, the eigenvalue $\nu_0$ satisfies 
$1-(\mu_s/\mu_t)\nu_0\tanh^{-1}(1/\nu_0)=0$. When $\mu_a\ll\mu_s$ as is typical in biological tissue, we have $1/\nu_0\approx\sqrt{3(1-\mu_s/\mu_t)}$ and
\be
\frac{\mu_t}{\nu_0}\approx\sqrt{3\mu_a\mu_t}.
\ee
Therefore we have $I\sim\exp(-z\sqrt{3\mu_a\mu_t+q_0^2})$.

In the general case of $g\neq0$, we can estimate $\nu_0$ using the fact that Case's eigenvalues are approximately obtained as eigenvalues of a tridiagonal matrix $B(M)$ (see (\ref{Bmatrix}) below). It is found \cite{Machida-etal09}
\be
\frac{\sqrt{1+\eta}}{\sqrt{3\frac{\mu_a}{\mu_t}\left(1-g\frac{\mu_s}{\mu_t}\right)}}\le\nu_0\le
\frac{1+\sqrt{\eta}}{\sqrt{3\frac{\mu_a}{\mu_t}\left(1-g\frac{\mu_s}{\mu_t}\right)}},
\ee
where
\be
\eta=\frac{4}{5}\frac{\mu_a}{\mu_a+\mu_s\left(1-g^2\right)}.
\ee

When $\mu_a$ is small, we have
\be
\frac{\mu_t}{\nu_0}\approx\sqrt{3\mu_a(\mu_t-g\mu_s)}\approx\sqrt{3\mu_a\mu_s'},
\ee
where $\mu_s'=(1-{\rm g})\mu_s$. Thus if $\mu_a$ is small we have in general
\be
I\sim e^{-z\sqrt{3\mu_a\mu_s'+q_0^2}}.
\label{diffdecay}
\ee

When $\mu_s'$ is large, the asymptotic decay of the diffusion equation is given by $\exp(-z\sqrt{\mu_{\rm eff}^2+q_0^2})$ in both of two diffusion approximations given in (\ref{case1u}) and (\ref{case2u}). Note that $\mu_{\rm eff}\approx\sqrt{3\mu_a\mu_s'}$ for small $\mu_a$. Since $\mu_t<\nu_0<\mu_{\rm eff}$ if $\eta>0$ is taken into account, the specific intensity decays slower than the prediction by the diffusion approximation.

If $q_0$ is large such that $\mu_t<\sqrt{\mu_{\rm eff}^2+q_0^2}$ or $\mu_*<\sqrt{\mu_{\rm eff}^2+q_0^2}$, the asymptotic decay from the diffusion approximation given in (\ref{case1u}) or (\ref{case2u}) becomes $\exp(-\mu_tz)$ or $\exp(-\mu_*z)$, and in either case the asymptotic behaviors of the RTE and diffusion equation are quite different.

\section{Spatial frequency domain imaging}
\label{sfdi}

Although different choices are possible we modulate the illuminating light in the $x$-direction and give the vector $\vv{q}_0$ as
\be
\vv{q}_0=(2\pi f,0)^T.
\label{q0f}
\ee

When the sample is illuminated by the source
\be
I_{\rm inc}(\vv{r},\uv)=\cos(2\pi fx)\delta(\uv-\hvv{z}).
\ee
The measured light is expressed as
\be
J_+(\vv{r}_d)=-A(q_0)\cos(2\pi fx_d),
\ee
where $x_d$ is the first component of $\vv{r}_d$. Here, $A(q_0)$ depends on $q_0=2\pi f$.

We use the amplitude $A(q_0)$ to reconstruct optical properties. Let $N_f$ be the number of spatial frequencies which are used for reconstruction. We have
\be
f=f_1,\dots,f_{N_f}.
\ee
Correspondingly, we write the forward data as $A(q_0^{(i)})$ ($i=1,\dots,N_f$). Let $\vv{y}\in\Rm^{N_f}$ be a vector defined as $\vv{y}=(A(q_0^{(i)})$. We can express $A_{\rm RTE}(q_0)$ as $A_{\rm RTE}(q_0^{(i)})$ ($i=1,\dots,N_f$). Similarly, $A_{\rm DA1}(q_0^{(i)})$ and $A_{\rm DA2}(q_0^{(i)})$ are introduced. Then computed values are stored in a vector $\vv{F}\in\Rm^{N_f}$.

Parameters $\mu_a,\mu_s'$ are determined by the Levenberg-Marquardt algorithm \cite{Levenberg44,Marquardt63}. To run the inversion algorithm with scaled variables $\xi_1,\xi_2$, we express the optical properties as \cite{Schweiger-Arridge99,Schweiger-Arridge-Nissila05}
\be
\mu_a=\mu_a^{(0)}e^{\xi_1},\quad\mu_s'={\mu_s'}^{(0)}e^{\xi_2},
\ee
where $\mu_a^{(0)},{\mu_s'}^{(0)}$ are initial guesses. That is, $\xi_1=\ln(\mu_a/\mu_a^{(0)})$, $\xi_2=\ln(\mu_s'/{\mu_s'}^{(0)})$.

We express $\vv{F}=\vv{F}(\vs{\xi})$, where $\vs{\xi}=(\xi_1,\xi_2)^T$. We wish to find
\be
\vs{\xi}_*=\argmin_{\vs{\xi}}\left|\vv{y}-\vv{F}(\vs{\xi})\right|^2.
\ee
Let $\vs{\xi}^k$ be the estimated solution at the $k$th iteration. Starting with the initial guess $\vs{\xi}^0=\vv{0}$, the solution is given by $\vs{\xi}_*=\lim_{k\to\infty}\vs{\xi}^k$. The FORTRAN library MINPACK \cite{More-Garbow-Hillstrom80} was used for the numerical calculation of the Levenberg-Marquardt method. Reconstructed values are obtained as
\be
\mu_a=\mu_a^{(0)}\exp(\xi_{*,1}),\quad\mu_s'=\mu_s{'(0)}\exp(\xi_{*,2}).
\ee
Below we will perform several parameter identifications using numerical phantoms.

\subsection{Thin slabs}
\label{3A}

\begin{figure}[htbp]
\centering
\includegraphics[width=0.3\textwidth]{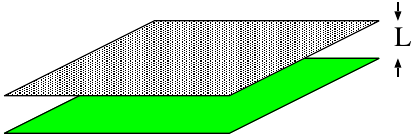}
\caption{
The thin slab.
}
\label{fig_slab}
\end{figure}

We perform parameter identification for numerical slab phantoms of size $90\,{\rm mm}\times 90\,{\rm mm}\times L$ as shown in Fig.~\ref{fig_slab}. The thickness $L$ changes from $1\,{\rm mm}$ to $10\,{\rm mm}$. The optical parameters of slabs are set to $\mu_a=0.02\,{\rm mm}^{-1}$, $\mu_s=10\,{\rm mm}^{-1}$, and ${\rm g}=0.9$. Moreover, $\mathfrak{n}=1$ (vacuum boundary condition). We suppose $\mu_a,\mu_s'$ are unknown. We set $N_f=2$, and $f_1=0.1$, $f_2=0.2$ (${\rm mm}^{-1}$).

As the forward data, the hemispheric flux is computed by Monte Carlo simulations and stored in $\vv{y}$. In each run, $10^8$ photons are launched. The vector $\vv{F}=(A_{\rm RTE}(q_0^{(i)}))$ ($i=1,2$) is computed from the RTE (see (\ref{Atheor2})). For the inverse problem of parameter identification, initial values are set to $(\mu_a^{(0)},\mu_s^{(0)})=(0.01\,{\rm mm}^{-1},10\,{\rm mm}^{-1})$.

\begin{figure}[htbp]
\centering
\includegraphics[width=0.45\textwidth]{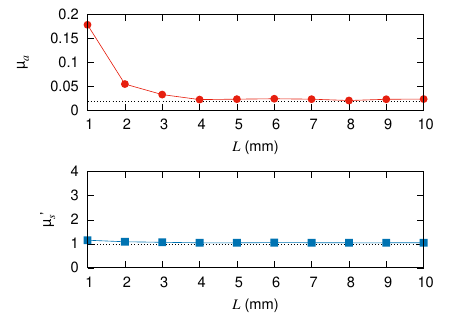}
\caption{
The upper panel shows estimated $\mu_a\,{\rm mm}^{-1}$ for slabs of thicknesses $1\,{\rm mm}$ through $10\,{\rm mm}$. The lower panel shows estimated $\mu_s'\,{\rm mm}^{-1}$ for the same slabs. Dotted lines show true values.
}
\label{figmc}
\end{figure}

Figure \ref{figmc} shows estimated $\mu_a$ and $\mu_s'$ for the slabs. Estimated values are $(\mu_a\,{\rm mm}^{-1}, \mu_s'\,{\rm mm}^{-1})=(0.18, 1.2)$, $(0.055, 1.1)$, $(0.033, 1.1)$, $(0.023, 1.0)$, $(0.023, 1.0)$, $(0.024, 1.1)$, $(0.023, 1.0)$, $(0.021, 1.0)$, $(0.023, 1.0)$, and $(0.024, 1.0)$ for $L=1\,{\rm mm}$, $2\,{\rm mm}$, $3\,{\rm mm}$, $4\,{\rm mm}$, $5\,{\rm mm}$, $6\,{\rm mm}$, $7\,{\rm mm}$, $8\,{\rm mm}$, $9\,{\rm mm}$, and $10\,{\rm mm}$, respectively. We see that for slabs of thickness larger than $4\,{\rm mm}$, optical properties are correctly obtained.

The behavior in Fig.~\ref{figmc} is implied in the decay in (\ref{diffdecay}) derived in Sec.~\ref{asym}. In the present situation we have
\be
I\sim e^{-2\pi fz}.
\ee
For $f=0.1\,{\rm mm}^{-1}$, we have $e^{-2\pi f\cdot3}=0.15$ and $e^{-2\pi f\cdot4}=0.08$. Hence the specific intensity is reduced by more than one tenth when the thickness of the slab is $4\,{\rm mm}$ or larger. Recently, an intensive study of Monte Carlo look-up tables was reported for relations between the penetration depths of photons and spatial frequencies \cite{Hayakawa-etal19}. Our conclusion in Fig.~\ref{figmc} is consistent with their results.

\subsection{Two-layer media}
\label{3B}

\begin{figure}[htbp]
\centering
\includegraphics[width=0.4\textwidth]{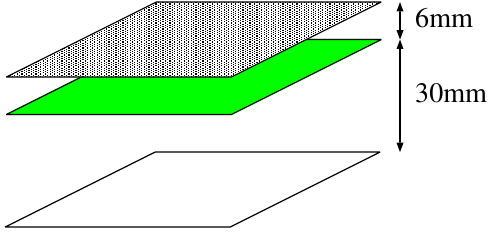}
\caption{
The two-layer medium.
}
\label{fig_two}
\end{figure}

Let us consider two-layer media, which have the top and bottom layers. As shown in Fig.~\ref{fig_two}, the thickness of the top layer is $6\,{\rm mm}$ and the bottom layer has the thickness of $30\,{\rm mm}$, which can be regarded as a semi-infinite medium. We set $N_f=2$, and $f_1=0.1$, $f_2=0.2$ (${\rm mm}^{-1}$). The scattering coefficient and anisotropic factor are fixed to $\mu_s=10\,{\rm mm}^{-1}$ and ${\rm g}=0.9$ in the entire medium. The refractive index is set to $\mathfrak{n}=1.4$. The absorption coefficient in the top layer is $\mu_a=0.02\,{\rm mm}^{-1}$. The absorption coefficient $\mu_a$ in the bottom layer takes values $0.01\,{\rm mm}^{-1}$ and $0.03\,{\rm mm}^{-1}$. We use $A_{\rm RTE}(q_0^{(i)})$ ($i=1,2$) for reconstruction. The forward data was computed by Monte Carlo simulations. Initial values were set to $(\mu_a^{(0)},\mu_s^{(0)})=(0.01\,{\rm mm}^{-1},10\,{\rm mm}^{-1})$.

Below, we present obtained $\mu_a,\mu_s'$ in the top layer when the absorption coefficient of the bottom layer is $0.01\,{\rm mm}^{-1}$ (Table \ref{fig_twoL6a}) and $0.03\,{\rm mm}^{-1}$ (Table \ref{fig_twoL6b}). To see how the estimated value is close to the true value, in Tables \ref{fig_twoL6a} and \ref{fig_twoL6b} we also give the relative error, which is defined as $|(\mbox{estimated value}-\mbox{true value})/\mbox{true value}|$. Numerical results show the reconstructed values are not affected by optical properties of the bottom layer.

\begin{table}[h!]
  \caption{Reconstructed $\mu_a$ and $\mu_s'$ (${\rm mm}^{-1}$) of the top layer of the two-layer medium by RTE for $\mu_a=0.01\,{\rm mm}^{-1}$ in the bottom layer.}
  \label{fig_twoL6a}
  \begin{center}
    \begin{tabular}{c|cc|cc}
    \hline
     & $\mu_a\,({\rm mm}^{-1})$ & error & $\mu_s'\,({\rm mm}^{-1})$ & error \\
    \hline
    true & 0.02  & - & 1.0 & - \\
    RTE  & 0.0079 & 0.60 & 1.1 & 0.14 \\
    \hline
    \end{tabular}
  \end{center}
\end{table}

\begin{table}[h!]
  \caption{Reconstructed $\mu_a$ and $\mu_s'$ (${\rm mm}^{-1}$) of the top layer of the two-layer medium by RTE for $\mu_a=0.03\,{\rm mm}^{-1}$ in the bottom layer.}
  \label{fig_twoL6b}
  \begin{center}
    \begin{tabular}{c|cc|cc}
    \hline
     & $\mu_a\,({\rm mm}^{-1})$ & error & $\mu_s'\,({\rm mm}^{-1})$ & error \\
    \hline
    true & 0.02 & - & 1.0 & - \\
    RTE  & 0.0087 & 0.57 & 1.1 & 0.13 \\
    \hline
    \end{tabular}
  \end{center}
\end{table}

\subsection{Three-layer media}
\label{3C}

\begin{figure}[htbp]
\centering
\includegraphics[width=0.4\textwidth]{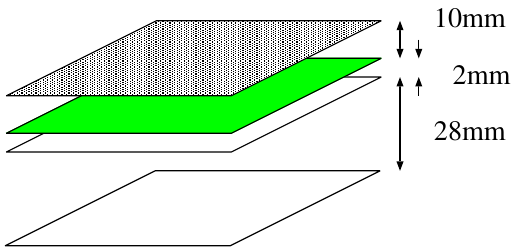}
\caption{
The three-layer medium.
}
\label{fig_three}
\end{figure}

Here we consider more complex media which have more than two layers \cite{Okada-Delpy03,Wang-etal19}. We reconstruct the optical properties of the top layer of a three layer medium shown in Fig.~\ref{fig_three}. Their optical properties are summarized in Table \ref{table_three}. The depths of layers are, from the top, $10\,{\rm mm}$, $2\,{\rm mm}$, and $28\,{\rm mm}$. From the top, $\mu_s=18$, $3$, and $21$ (${\rm mm}^{-1}$). In the 2nd layer, $\mu_a=0.004\,{\rm mm}^{-1}$. The absorption coefficients of the 1st and 3rd layers vary from $0.01\,{\rm mm}^{-1}$ to $0.03\,{\rm mm}^{-1}$. In addition, ${\rm g}=0.9$ and $\mathfrak{n}=1.4$. We set $N_f=2$, and $f_1=1/15$, $f_2=1/10$ (${\rm mm}^{-1}$). Starting the Levenberg-Marquardt algorithm with initial values $(\mu_a^{(0)},\mu_s^{(0)})=(0.01\,{\rm mm}^{-1},10\,{\rm mm}^{-1})$, we obtain $\mu_a,\mu_s'$ after about ten iterations.

\begin{table}[h!]
  \caption{Three-layer model. In the top layer, $\mu_{a,1}=0.01$, $0.02$, or $0.03$ (${\rm mm}^{-1}$). In the bottom layer, $\mu_{a,3}=0.01$, $0.02$, or $0.03$ (${\rm mm}^{-1}$).}
  \label{table_three}
  \begin{center}
    \begin{tabular}{c|ccc}
    \hline
     & $\mu_a\,({\rm mm}^{-1})$ & $\mu_s\,({\rm mm}^{-1})$ & $d\,({\rm mm})$ \\
    \hline
    1st layer & $\mu_{a,1}$  & 18 & 10 \\
    2nd layer & 0.004 & \phantom{0}3 & \phantom{0}2 \\
    3rd layer & $\mu_{a,3}$  & 21 & 28 \\
    \hline
    \end{tabular}
  \end{center}
\end{table}

For comparison, reconstructions by the diffusion approximation are also obtained. That is, $A_{\rm DA1}$ in (\ref{JpADA1}) and $A_{\rm DA2}$ in (\ref{JpADA2}) are used in addition to $A_{\rm RTE}$ in (\ref{JplusA}). The forward data were prepared by Monte Carlo simulations.

\begin{table}[h!]
  \caption{Reconstructed $\mu_a$ and $\mu_s'$ (${\rm mm}^{-1}$) by RTE, DA1, and DA2 when $\mu_a=0.01\,{\rm mm}^{-1}$ in the top layer and $\mu_a=0.01\,{\rm mm}^{-1}$ in the bottom layer.}
  \label{fig_top001a}
  \begin{center}
    \begin{tabular}{c|cc|cc}
    \hline
     & $\mu_a\,({\rm mm}^{-1})$ & error & $\mu_s'\,({\rm mm}^{-1})$ & error \\
    \hline
    true & 0.01 & - & 1.8 & - \\
    RTE  & 0.0084 & 0.16 & 1.8 & 0.021 \\
    DA1  & 0.0060 & 0.40 & 1.8 & 0.018 \\
    DA2  & 0.015  & 0.53 & 1.5 & 0.18 \\
    \hline
    \end{tabular}
  \end{center}
\end{table}

\begin{table}[h!]
  \caption{Reconstructed $\mu_a$ and $\mu_s'$ (${\rm mm}^{-1}$) by RTE, DA1, and DA2 when $\mu_a=0.01\,{\rm mm}^{-1}$ in the top layer and $\mu_a=0.02\,{\rm mm}^{-1}$ in the bottom layer.}
  \label{fig_top001b}
  \begin{center}
    \begin{tabular}{c|cc|cc}
    \hline
     & $\mu_a\,({\rm mm}^{-1})$ & error & $\mu_s'\,({\rm mm}^{-1})$ & error \\
    \hline
    true & 0.01 & - & 1.8 & - \\
    RTE  & 0.0085 & 0.15 & 1.8 & 0.021 \\
    DA1  & 0.0060 & 0.40 & 1.8 & 0.018 \\
    DA2  & 0.015  & 0.53 & 1.5 & 0.18 \\
    \hline
    \end{tabular}
  \end{center}
\end{table}

\begin{table}[h!]
  \caption{Reconstructed $\mu_a$ and $\mu_s'$ (${\rm mm}^{-1}$) by RTE, DA1, and DA2 when $\mu_a=0.01\,{\rm mm}^{-1}$ in the top layer and $\mu_a=0.03\,{\rm mm}^{-1}$ in the bottom layer.}
  \label{fig_top001c}
  \begin{center}
    \begin{tabular}{c|cc|cc}
    \hline
     & $\mu_a\,({\rm mm}^{-1})$ & error & $\mu_s'\,({\rm mm}^{-1})$ & error \\
    \hline
    true & 0.01 & - & 1.8 & - \\
    RTE  & 0.0085 & 0.15 & 1.8 & 0.022 \\
    DA1  & 0.0060 & 0.40 & 1.8 & 0.018 \\
    DA2  & 0.015  & 0.54 & 1.5 & 0.18 \\
    \hline
    \end{tabular}
  \end{center}
\end{table}

\begin{table}[h!]
  \caption{Reconstructed $\mu_a$ and $\mu_s'$ (${\rm mm}^{-1}$) by RTE, DA1, and DA2 when $\mu_a=0.02\,{\rm mm}^{-1}$ in the top layer and $\mu_a=0.01\,{\rm mm}^{-1}$ in the bottom layer.}
  \label{fig_top002a}
  \begin{center}
    \begin{tabular}{c|cc|cc}
    \hline
     & $\mu_a\,({\rm mm}^{-1})$ & error & $\mu_s'\,({\rm mm}^{-1})$ & error \\
    \hline
    true & 0.02 & - & 1.8 & - \\
    RTE  & 0.016 & 0.20 & 1.8 & 0.0015 \\
    DA1  & 0.012 & 0.39 & 1.9 & 0.053 \\
    DA2  & 0.026 & 0.31 & 1.5 & 0.19 \\
    \hline
    \end{tabular}
  \end{center}
\end{table}

\begin{table}[h!]
  \caption{Reconstructed $\mu_a$ and $\mu_s'$ (${\rm mm}^{-1}$) by RTE, DA1, and DA2 when $\mu_a=0.02\,{\rm mm}^{-1}$ in the top layer and $\mu_a=0.02\,{\rm mm}^{-1}$ in the bottom layer.}
  \label{fig_top002b}
  \begin{center}
    \begin{tabular}{c|cc|cc}
    \hline
     & $\mu_a\,({\rm mm}^{-1})$ & error & $\mu_s'\,({\rm mm}^{-1})$ & error \\
    \hline
    true & 0.02 & - & 1.8 & - \\
    RTE  & 0.016 & 0.20 & 1.8 & 0.0013 \\
    DA1  & 0.012 & 0.39 & 1.9 & 0.053 \\
    DA2  & 0.026 & 0.31 & 1.5 & 0.19 \\
    \hline
    \end{tabular}
  \end{center}
\end{table}

\begin{table}[h!]
  \caption{Reconstructed $\mu_a$ and $\mu_s'$ (${\rm mm}^{-1}$) by RTE, DA1, and DA2 when $\mu_a=0.02\,{\rm mm}^{-1}$ in the top layer and $\mu_a=0.03\,{\rm mm}^{-1}$ in the bottom layer.}
  \label{fig_top002c}
  \begin{center}
    \begin{tabular}{c|cc|cc}
    \hline
     & $\mu_a\,({\rm mm}^{-1})$ & error & $\mu_s'\,({\rm mm}^{-1})$ & error \\
    \hline
    true & 0.02 & - & 1.8 & - \\
    RTE  & 0.016 & 0.20 & 1.8 & 0.0012 \\
    DA1  & 0.012 & 0.39 & 1.9 & 0.053 \\
    DA2  & 0.026 & 0.32 & 1.5 & 0.19 \\
    \hline
    \end{tabular}
  \end{center}
\end{table}

\begin{table}[h!]
  \caption{Reconstructed $\mu_a$ and $\mu_s'$ (${\rm mm}^{-1}$) by RTE, DA1, and DA2 when $\mu_a=0.03\,{\rm mm}^{-1}$ in the top layer and $\mu_a=0.01\,{\rm mm}^{-1}$ in the bottom layer.}
  \label{fig_top003a}
  \begin{center}
    \begin{tabular}{c|cc|cc}
    \hline
     & $\mu_a\,({\rm mm}^{-1})$ & error & $\mu_s'\,({\rm mm}^{-1})$ & error \\
    \hline
    true & 0.03 & - & 1.8 & - \\
    RTE  & 0.023 & 0.25 & 1.8 & 0.026 \\
    DA1  & 0.017 & 0.42 & 2.0 & 0.088 \\
    DA2  & 0.036 & 0.21 & 1.5 & 0.18 \\
    \hline
    \end{tabular}
  \end{center}
\end{table}

\begin{table}[h!]
  \caption{Reconstructed $\mu_a$ and $\mu_s'$ (${\rm mm}^{-1}$) by RTE, DA1, and DA2 when $\mu_a=0.03\,{\rm mm}^{-1}$ in the top layer and $\mu_a=0.02\,{\rm mm}^{-1}$ in the bottom layer.}
  \label{fig_top003b}
  \begin{center}
    \begin{tabular}{c|cc|cc}
    \hline
     & $\mu_a\,({\rm mm}^{-1})$ & error & $\mu_s'\,({\rm mm}^{-1})$ & error \\
    \hline
    true & 0.03 & - & 1.8 & - \\
    RTE  & 0.023 & 0.25 & 1.8 & 0.026 \\
    DA1  & 0.017 & 0.42 & 2.0 & 0.088 \\
    DA2  & 0.036 & 0.21 & 1.5 & 0.18 \\
    \hline
    \end{tabular}
  \end{center}
\end{table}

\begin{table}[h!]
  \caption{Reconstructed $\mu_a$ and $\mu_s'$ (${\rm mm}^{-1}$) by RTE, DA1, and DA2 when $\mu_a=0.03\,{\rm mm}^{-1}$ in the top layer and $\mu_a=0.03\,{\rm mm}^{-1}$ in the bottom layer.}
  \label{fig_top003c}
  \begin{center}
    \begin{tabular}{c|cc|cc}
    \hline
     & $\mu_a\,({\rm mm}^{-1})$ & error & $\mu_s'\,({\rm mm}^{-1})$ & error \\
    \hline
    true & 0.03 & - & 1.8 & - \\
    RTE  & 0.023 & 0.25 & 1.8 & 0.026 \\
    DA1  & 0.017 & 0.42 & 2.0 & 0.088 \\
    DA2  & 0.036 & 0.21 & 1.5 & 0.18 \\
    \hline
    \end{tabular}
  \end{center}
\end{table}

Tables \ref{fig_top001a} through \ref{fig_top003c} show reconstructed $\mu_a,\mu_s'$ when the true value in the top layer is $\mu_a=0.01\,{\rm mm}^{-1}$, $0.02\,{\rm mm}^{-1}$, and $0.03\,{\rm mm}^{-1}$, respectively. In each Table, relative errors are also shown. The true value of $\mu_s'$ in the top layer is fixed to $1.8\,{\rm mm}^{-1}$ as shown in Table \ref{table_three}. We see that reconstructed values by DA1 are closer to those by RTE compared with reconstructed values by DA2. The absorption coefficient of the bottom layer is $\mu_a=0.01\,{\rm mm}^{-1}$ in Tables \ref{fig_top001a}, \ref{fig_top002a}, \ref{fig_top003a}, $\mu_a=0.02\,{\rm mm}^{-1}$ in Tables \ref{fig_top001b}, \ref{fig_top002b}, \ref{fig_top003b}, and $\mu_a=0.03\,{\rm mm}^{-1}$ in Tables \ref{fig_top001c}, \ref{fig_top002c}, \ref{fig_top003c}. In all cases, reconstructed values are not affected by the third layer. 

\section{Solid phantom}
\label{experiments}

\begin{figure}[htbp]
\centering
\includegraphics[width=0.33\textwidth]{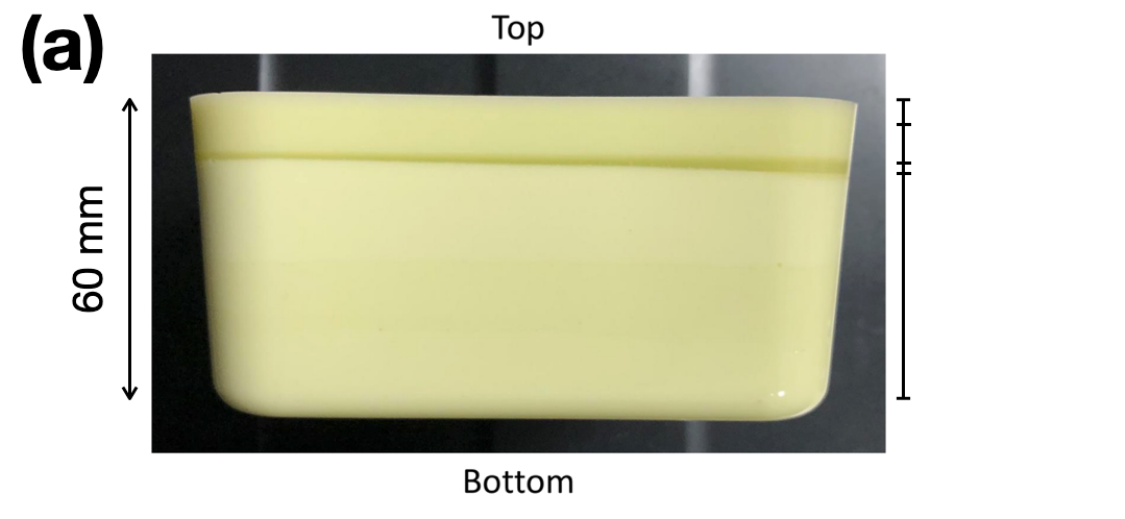}
\includegraphics[width=0.33\textwidth]{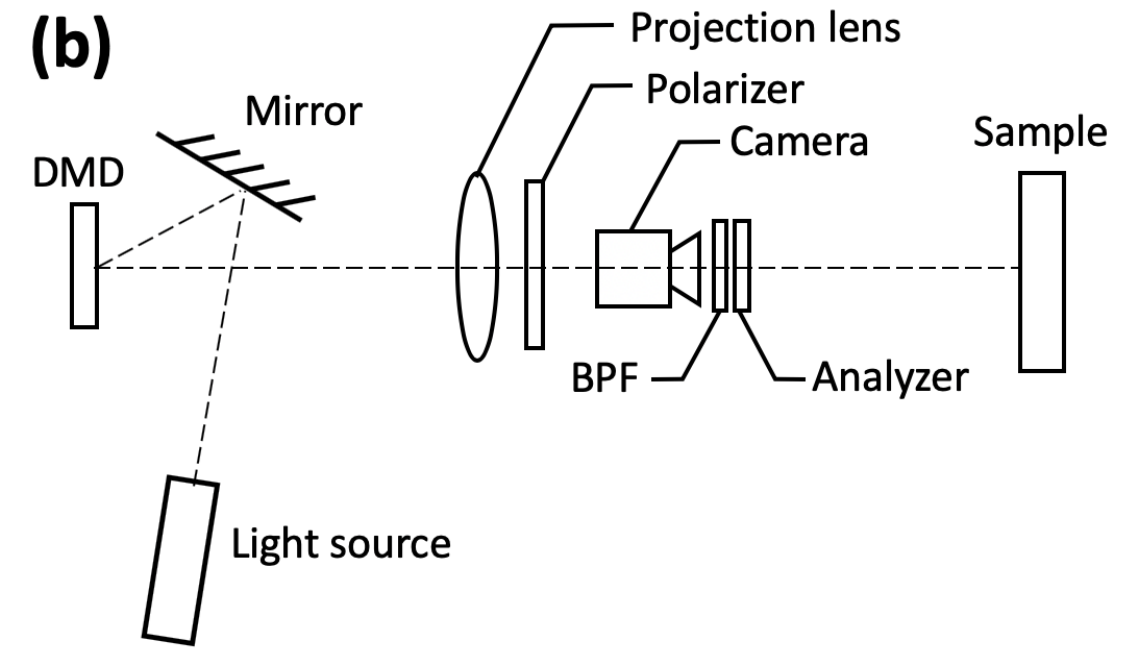}
\includegraphics[width=0.33\textwidth]{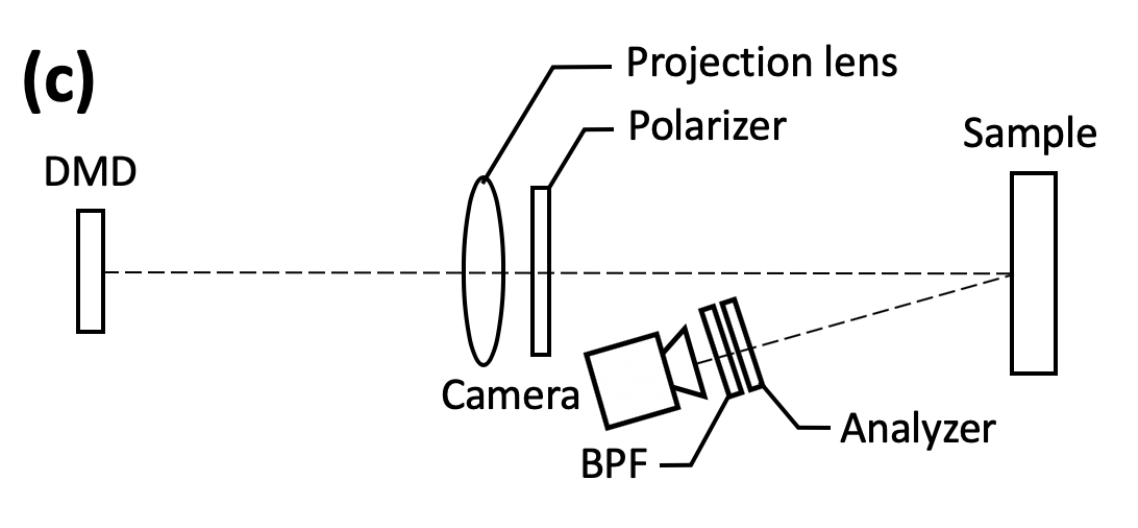}
\caption{
(a) Four-layer solid phantom. The width of the top layer is about $4\,{\rm mm}$, and the width of the bottom layer is about $47\,{\rm mm}$. Measurement setup: (b) top view and (c) side view. DMD, digital micro-mirror device. BPF, band-pass filter.
}
\label{figpic}
\end{figure}

Figure \ref{figpic}(a) shows a solid phantom made of epoxy resin. The refractive index of the phantom is $1.58$. The phantom has a four-layer structure and the width of the top layer is about $4\,{\rm mm}$. When the phantom was made, optical properties of the top layer were aimed at $\mu_a=0.0165\,{\rm mm}^{-1}$ and $\mu_s'=1.3\,{\rm mm}^{-1}$. The setup of the measurement system is shown in Figs.~\ref{figpic}(b) and (c). As a near-infrared light source, a broadband halogen fiber optic illuminator (Thorlabs, OSL2) with an enhanced infrared replacement bulb (Thorlabs, OSL2BIR) was used. The projected patterns were displayed on a digital micro-mirror device (DMD) module (Keynote Photonics, LC4500 NIR controller). The image was formed on the sample plane through a visible-near-infrared lens (Schneider, large format F-mount lens, focal length of 28mm, F/2.8). To capture reflection images, a camera with enhanced near-infrared sensitivity (Ximea, MQ-13RG-E2, 1280x1024 pixels, monochrome) with a visible-near-infrared lens (Edmund Optics, C series VIS-NIR fixed focal length lens, focal length of 16mm, F/1.6) were used. The polarizer and analyzer placed in the crossed nicols configuration were inserted into the measurement system to remove the specular reflection from the sample. A band-pass filter (Edmund Optics, hard coated OD 4.0 25nm bandpass filter, center wavelength of 800nm, FWHM of 25nm) was placed in front of the camera to extract the 800nm wavelength. The top layer of the phantom was illuminated by the spatially modulated light and the reflected light was detected by the camera.

The source term in experiments is given by
\be
I_{\rm inc}(\vv{r},\uv)=\frac{S_0}{2}\left[1+\cos\left(2\pi fx+\alpha\right)\right]\delta(\uv-\hvv{z}),
\ee
where $\alpha=2\pi p/3$ ($p=0,1,2$). We set $N_f=2$, and $f_1=0.1$, $f_2=0.2$ (${\rm mm}^{-1}$). We have
\be
J_+(\vv{r}_d;\alpha)=\frac{S_0}{2}\left(
M_{\rm DC}+M_{\rm AC}\cos(2\pi fx_d+\alpha)\right),
\ee
where $M_{\rm DC},M_{\rm AC}$ depend on $\vv{r}_d$ and $f$ in general. Let us write $J_+^{(p)}=J_+(\vv{r}_d;2\pi p/3)$. By a straightforward calculation, we have
\be
\begin{aligned}
&
\left(M_{\rm AC}\right)^2=
\frac{2}{9}
\\
&\times
\left[\left(J_+^{(0)}-J_+^{(1)}\right)^2+
\left(J_+^{(1)}-J_+^{(2)}\right)^2+\left(J_+^{(2)}-J_+^{(0)}\right)^2\right].
\end{aligned}
\ee
Moreover we can write \cite{Cuccia-etal09}
\be
M_{\rm AC}=M^{\rm exp}(\vv{r}_d,f)A^{\rm exp}(q_0^{(i)}),
\ee
where $M^{\rm exp}$ ($>0$) is a constant determined by the optical system and $q_0^{(i)}=f_i$ ($i=1,2$).

Let us consider $M_{\rm AC,\,ref}$ of a reference medium, for which $A_{\rm ref}^{\rm exp}(q_0)$ can be numerically computed. Then we have \cite{Cuccia-etal09}
\be
A^{\rm exp}(q_0)=
\left\langle\frac{M_{\rm AC}}{M_{\rm AC,\,ref}}\right\rangle
A_{\rm ref}^{\rm exp}(q_0),
\label{Anexp}
\ee
where $\langle\cdot\rangle$ means the average in space.

We use the bottom layer of the solid phantom as the reference medium. Since the fourth layer of the phantom has the thickness $47\,{\rm mm}$, it can be regarded as the half space even for the time-resolved measurements in which two optical fibers are vertically attached on the bottom side of the phantom. From time-resolved measurements by TRS-80 (Hamamatsu Photonics), we found $\mu_a=0.010\,{\rm mm}^{-1}$ and $\mu_s'=1.4\,{\rm mm}^{-1}$ for the bottom layer. With these optical parameters, $A_{\rm ref}^{\rm exp}(q_0^{(i)})$ can be computed from the RTE as $A_{\rm RTE}(q_0^{(i)})$. Thus, $A^{\rm exp}(q_0^{(i)})$ in (\ref{Anexp}) are prepared and stored in a vector $\vv{y}\in\Rm^{N_f}$.

The Levenberg-Marquardt algorithm was run with initial values $\mu_a^{(0)}=0.01\,{\rm mm}^{-1}$ and ${\mu_s}^{(0)}=10\,{\rm mm}^{-1}$. We obtain
\be
\mu_a=0.016\,{\rm mm}^{-1},\quad\mu_s'=1.0\,{\rm mm}^{-1}.
\label{obtainedparam}
\ee
Other choices of initial guess, for example $\mu_a^{(0)}=0.02\,{\rm mm}^{-1}$ and ${\mu_s'}^{(0)}=2\,{\rm mm}^{-1}$, give the same $\mu_a,{\mu_s'}$ given in (\ref{obtainedparam}). The computation time was less than $1$ sec with a laptop computer (MacBook Pro with $2.3$ GHz Intel Core i5 and $8$ GB memory).

\section{Discussion and conclusions}
\label{concl}

Taking advantage of the fact that near-infrared light decays rapidly for nonzero spatial frequencies, in this paper, we estimated optical properties of the top layer of the layered phantom. Indeed, SFDI has been used for the parameter identification of the top layer. Various numerical experiments developed in the present paper confirm that results by SFDI are not affected by deeper layers. Figure \ref{figmc} indicates that the necessary width of the top layer is $4\,{\rm mm}$ (when $1/\mu_s'=1\,{\rm mm}$). Our investigation in the asymptotic limit provides a theoretical reason for the results of numerical experiments. Using the numerical algorithm based on the method of rotated reference frames, optical properties of the top layer of a layered solid phantom were also determined.

Our numerical results show that $\mu_s'$ is more accurately reconstructed than $\mu_a$. One of the reasons of this inaccuracy is the ill-posedness of the parameter identification. The future study on regularization might improve the results.

When two optical fibers are attached to biological tissue in the direction perpendicular to its surface with the separation a few centimeters, the detected reflected near-infrared light contains photons from depths more than a centimeter. Although this very feature makes it possible to study the function of the human brain through the neurovascular coupling \cite{Ferrari-Quaresim12}, the detected light is affected by optical properties of different layers. In particular, the signal from the brain is affected by skin blood flow in the scalp \cite{Hoshi16}. In this conventional way, it is not possible to extract only information on shallow regions. Photons which travel deep inside biological tissue can be excluded if the separation of two optical fibers is reduced. However, then measurements have to be conducted in a tiny space and other difficulties related to measurements arise (see \cite{Kohno-Hoshi16} and references therein). The SFDI measurement setup described in Fig.~\ref{figpic}(a) is free from such difficulties.

For our approach to work, the top layer of a layered random medium has to be regarded as a semi-infinite medium. The necessary width of the top layer depends on spatial frequency. This can be checked by a test parameter identification for numerical phantoms.

When $\mu_a$ is not small, the decay of the specific intensity deviates from the diffusive decay and is given by (\ref{mainthm}) because then $\mu_t/\nu_0$ can not be approximated by $\sqrt{3\mu_a\mu_s'}$. Moreover if the depth is not large compared to $\nu_0/\mu_t$ nor $1/q_0$, not only $I_0$ but other modes contribute, and the decay is given by the superposition of different decays shown in (\ref{decaymodes}).

\section*{Funding.}

The authors appreciate HUSM Grant-in-Aid funded by Hamamatsu University School of Medicine. MM acknowledges support from JSPS KAKENHI Grant No.~17K05572, 17H02081, 18K03438. YH acknowledges support from JSPS KAKENHI Grant No.~17H02081. KK acknowledges support from JSPS KAKENHI Grant No.~16K04985, 17H06102, 18H01497, 18H05240.

\section*{Acknowledgments.}

The Monte Carlo simulation in Sec.~\ref{sfdi}.\ref{3A} was carried out using the package MC written by Vadim A. Markel (\texttt{http://whale.seas.upenn.edu/vmarkel/CODES/MC.html}). The Monte Carlo eXtreme (MCX) (\texttt{http://mcx.space/}) was used for Monte Carlo simulations in Secs.~\ref{sfdi}.\ref{3B} and \ref{3C}. The solid phantom was provided by Yukari Tanikawa.

\appendix

\section{Diffusion approximation}
\label{rte}

We begin by decomposing $I$ into the following two terms:
\be
I(\vv{r},\uv)=I_b(\vv{r},\uv)+I_s(\vv{r},\uv).
\label{decom}
\ee
The ballistic term $I_b$ and scattering term $I_s$ satisfy
\be
\left\{\begin{aligned}
\left(\uv\cdot\nabla+\mu_a+\mu_s\right)I_b(\vv{r},\uv)=0,
&\quad (\vv{r},\uv)\in\Omega\times\Sm^2,
\\
I_b(\vv{r},\uv)=I_{\rm inc}(\vv{r},\uv),
&\quad (\vv{r},\uv)\in\Gamma_-,
\end{aligned}\right.
\label{rteIb}
\ee
and
\be
\left\{\begin{aligned}
&
\left(\uv\cdot\nabla+\mu_a+\mu_s\right)I_s(\vv{r},\uv)=
\mu_s\int_{\Sm^2}p(\uv,\uv')I_s(\vv{r},\uv')\,d\uv'
\\
&\qquad+
S(\vv{r},\uv),
\quad (\vv{r},\uv)\in\Omega\times\Sm^2,
\\
&
I_s(\vv{r},\uv)=
R_{\mathfrak{n}}(\uv\cdot\hvv{z})I_s(\vv{r},\uv_R),
\quad (\vv{r},\uv)\in\Gamma_-.
\end{aligned}\right.
\label{rteIs}
\ee
Here the source term for $I_s$ is given by
\be
S(\vv{r},\uv)=\mu_s\int_{\Sm^2}p(\uv,\uv')I_b(\vv{r},\uv')\,d\uv'.
\ee
Since
\be
I_b(\vv{r},\uv)=e^{-\mu_tz}e^{i\vv{q}_0\cdot\vs{\rho}}\delta(\uv-\hvv{z}),
\ee
we have
\be
S(\vv{r},\uv)=\mu_se^{i\vv{q}_0\cdot\vs{\rho}}e^{-\mu_tz}p(\uv,\hvv{z}).
\ee

Suppose that $I_s$ weakly depends on $\uv$ and can be written as
\be
I_s(\vv{r},\uv)=\frac{1}{4\pi}u(\vv{r})+\frac{3}{4\pi}\vv{J}(\vv{r})\cdot\uv,
\ee
where
\be
u(\vv{r})=\int_{\Sm^2}I_s(\vv{r},\uv)\,d\uv,\quad
\vv{J}(\vv{r})=\int_{\Sm^2}\uv I_s(\vv{r},\uv)\,d\uv.
\ee
Let us write
\be
u(\vv{r})=v_{\rm DA1}(z)e^{i\vv{q}_0\cdot\vs{\rho}}.
\label{uandv}
\ee
Then from (\ref{rteIs}), we obtain
\be
\frac{\pp^2}{\pp z^2}v_{\rm DA1}-\left(\mu_{\rm eff}^2+q_0^2\right)v_{\rm DA1}=-Be^{-\mu_tz},
\ee
where
\be
B=3\mu_*\mu_s\left(1+{\rm g}\frac{\mu_t}{\mu_*}\right).
\label{DA1B}
\ee
We note that
\be
\vv{J}(\vv{r})=-\frac{1}{3\mu_*}\nabla u(\vv{r})+\frac{g\mu_s}{\mu_*}e^{i\vv{q}_0\cdot\vs{\rho}}e^{-\mu_tz}\hvv{z}.
\ee
We obtain
\be
v_{\rm DA1}(z)=\frac{-B}{\mu_t^2-\mu_{\rm eff}^2-q_0^2}e^{-\mu_tz}
+C_1e^{-\sqrt{\mu_{\rm eff}^2+q_0^2}z},
\label{vDA1BC}
\ee
where $C_1$ is determined from the boundary condition. If we assume the diffuse boundary condition such that
\be
-D_0\frac{\pp u}{\pp z}+\frac{1}{\zeta}u=0,
\ee
where $D_0=1/(3\mu_*)$ and $\zeta=2(1+r_d)/(1-r_d)$ with
\be
r_d=-1.4399\mathfrak{n}^{-2}+0.7099\mathfrak{n}^{-1}+0.6681+0.0636\mathfrak{n},
\ee
we obtain
\be
C_1=\frac{B}{\mu_t^2-\mu_{\rm eff}^2-q_0^2}\frac{\zeta D_0\mu_t+1}{\zeta D_0\sqrt{\mu_{\rm eff}^2+q_0^2}+1}.
\label{DA1C}
\ee
On the boundary at $\vv{r}_d$, we have
\be
u(\vv{r}_d)=
\frac{3\mu_*\mu_s(1+{\rm g}\mu_t/\mu_*)e^{i\vv{q}_0\cdot\vs{\rho}}}{\left(\sqrt{\mu_{\rm eff}^2+q_0^2}+\mu_t\right)\left(\sqrt{\mu_{\rm eff}^2+q_0^2}+3\mu_*/\zeta\right)}.
\label{urd1}
\ee

We obtain
\be
\begin{aligned}
J_+(\vv{r}_d)
&=
\int_{\Sm^2_-}(\cos\vth)\left(\frac{1}{4\pi}u(\vv{r}_d)+\frac{3}{4\pi}\vv{J}(\vv{r}_d)\cdot\uv\right)\,d\uv
\\
&=
-\frac{1}{4}u(\vv{r}_d)+\frac{1}{2}J_z(\vv{r}_d)
\\
&=
-A_{\rm DA1}(q_0)e^{i\vv{q}_0\cdot\vs{\rho}},
\end{aligned}
\label{da:naive2}
\ee
where $u(\vv{r}_d)$ is given in (\ref{urd1}). Here we introduced
\be
A_{\rm DA1}(q_0)=
\left(\frac{1}{4}+\frac{1}{2\zeta}\right)v_{\rm DA1}(0)-\frac{{\rm g}\mu_s}{2\mu_*}.
\label{JpADA1}
\ee

Next, instead of (\ref{rteIb}), let us introduce the ballistic term as 
$(\uv\cdot\nabla+\mu_*)I_0=0$, we obtain
\be
I_0(\vv{r},\uv)=e^{-\mu_*z}e^{i\vv{q}_0\cdot\vs{\rho}}\delta(\uv-\hvv{z}),
\ee
and
\be
\begin{aligned}
S(\vv{r},\uv)
&=
(\mu_*-\mu_t)I_0(\vv{r},\uv)+\mu_s\int_{\Sm^2}p(\uv,\uv')I_0(\vv{r},\uv')\,d\uv'
\\
&=
\left(-\mu_s{\rm g}\delta(\uv-\hvv{z})+\mu_sp(\uv,\hvv{z})\right)e^{-\mu_*z}e^{i\vv{q}_0\cdot\vs{\rho}}.
\end{aligned}
\ee
We can write
\be
u(\vv{r})=v_{\rm DA2}(z)e^{i\vv{q}_0\cdot\vs{\rho}}.
\ee
After similar calculations, we obtain
\be
\vv{J}(\vv{r})=-D_0\nabla u(\vv{r}),
\ee
and
\be
\frac{\pp^2}{\pp z^2}v_{\rm DA2}-\left(\mu_{\rm eff}^2+q_0^2\right)v_{\rm DA2}=-3\mu_*\mu_s'e^{-\mu_*z}.
\ee
Hence,
\be
v_{\rm DA2}(z)=\frac{-3\mu_*\mu_s'}{\mu_*^2-\mu_{\rm eff}^2-q_0^2}e^{-\mu_*z}
+C_2e^{-\sqrt{\mu_{\rm eff}^2+q_0^2}z},
\label{vDA2BC}
\ee
where
\be
C_2=\frac{3\mu_*\mu_s'}{\mu_*^2-\mu_{\rm eff}^2-q_0^2}\frac{\zeta D_0\mu_*+1}{\zeta D_0\sqrt{\mu_{\rm eff}^2+q_0^2}+1}.
\label{DA2C}
\ee
We obtain
\be
u(\vv{r}_d)=
\frac{3\mu_*\mu_s'e^{i\vv{q}_0\cdot\vs{\rho}}}{\left(\sqrt{\mu_{\rm eff}^2+q_0^2}+\mu_*\right)\left(\sqrt{\mu_{\rm eff}^2+q_0^2}+3\mu_*/\zeta\right)}.
\label{urd2}
\ee
In this case we obtain
\be
J_+(\vv{r}_d)=-A_{\rm DA2}(q_0)e^{i\vv{q}_0\cdot\vs{\rho}},
\label{da:ellsta}
\ee
where
\be
A_{\rm DA2}(q_0)=\left(\frac{1}{4}+\frac{1}{2\zeta}\right)v_{\rm DA2}(0).
\label{JpADA2}
\ee
The difference between (\ref{urd1}) and (\ref{urd2}) is small when $g$ is small. In this paper we use (\ref{urd1}), for which the decomposition is compatible with the ballistic subtraction developed in Appendix \ref{mrrfmain}.

\section{The method of rotated reference frames}
\label{mrrfmain}

For the method of rotated reference frames in the half space, the subtraction of the ballistic term was considered \cite{Liemert-Kienle13}. Here, we will compute the hemispheric flux $J_+$ following \cite{Liemert-Kienle13} with a spatially oscillating source term.

\subsection*{B. 1. Preliminary}

We begin with the one-dimensional RTE:
\be
\left\{\begin{aligned}
\left(\cos\vth\frac{\pp}{\pp z}+\mu_t\right)I_1(z,\uv)=
\mu_s\int_{\Sm^2}p(\uv,\uv')I_1(z,\uv')\,d\uv',
\\
(z,\uv)\in(0,\infty)\times\Sm^2,
\\
I_1(z,\uv)=R_{\mathfrak{n}}(\mu)I_1(z,\uv_R)+g_1(z,\uv),
\\
\uv\in\Sm^2_+\quad\mbox{at}\;z=0
\end{aligned}\right.
\ee
with a source term $g_1(z,\uv)$. In one-dimensional transport theory, it is known that the solution $I_1$ is expressed as \cite{Case-Zweifel}
\be
\begin{aligned}
I_1(z,\uv)
&=
\sum_{M=-l_{\rm max}}^{l_{\rm max}}
\Biggl[\sum_{j=1}^{J^M}
\tilde{a}_j^M\Phi_{\nu_j(M)}^M(\uv)e^{-\mu_tz/\nu_j(M)}
\\
&+
\int_0^1
\tilde{a}^M(\nu)\Phi_{\nu}^M(\uv)e^{-\mu_tz/\nu}\,d\nu
\Biggr],
\end{aligned}
\ee
where coefficients $\tilde{a}_j^M$, $\tilde{a}^M(\nu)$ are determined from the boundary condition. Here, $\Phi_{\nu}^M(\uv)$ ($\nu=\nu_j(M)$ or $\nu\in(0,1)$) is given by \cite{Case-Zweifel,Case60,McCormick-Kuscer66,Mika61}
\be
\Phi_{\nu}^M(\uv)=
\phi^M(\nu,\cos\vth)\left(1-\cos^2\vth\right)^{|M|/2}e^{iM\va},
\ee
where $\phi^M(\nu,\cos\vth)$ is called Case's singular eigenfunction.

Let us introduce $\sigma_l>0$ as
\be
\sigma_l=\mu_t-\mu_s{\rm g}^l=\mu_a+(1-{\rm g}^l)\mu_s
\ee
and introduce $b_l(m)$ as
\be
b_l(m)=\sqrt{(l^2-m^2)/((4l^2-1)\sigma_l\sigma_{l-1})}.
\ee
We consider the normalized Chandrasekhar polynomial $g_l^m(x)$, which satisfies the following three-term recurrence relation \cite{Garcia-Siewert89,Garcia-Siewert90}.
\be
\begin{aligned}
\frac{x}{\mu_t}\sqrt{(2l+1)\sigma_l}g_l^m(x)
&=
b_{l+1}(m)\sqrt{(2l+3)\sigma_{l+1}}g_{l+1}^m(x)
\\
&+
b_l(m)\sqrt{(2l-1)\sigma_{l-1}}g_{l-1}^m(x)
\end{aligned}
\ee
for $l>m$ and $m\ge0$. We have
\be
\begin{aligned}
&
g_m^m(x)=\frac{(2m-1)!!}{\sqrt{(2m)!}}=\frac{\sqrt{(2m)!}}{2^mm!},
\\
&
g_l^{-m}(x)=(-1)^mg_l^m(x),
\\
&
g_l^m(-x)=(-1)^{l+m}g_l^m(x).
\end{aligned}
\ee
Now, eigenvalues $\nu_j(M)$ are zeros of $g_l^M$ as $l\to\infty$ \cite{Garcia-Siewert82}. In Sec.~\ref{generalsol} we numerically obtain $\nu_j(M)$ and $\nu\in(0,1)$ as eigenvalues of a tridiagonal matrix.

Let $f(\uv)$ be a function of $\uv\in\Sm^2$ which can be expressed as $f(\uv)=\sum_{l=0}^{\infty}\sum_{m=-l}^lf_{lm}Y_{lm}(\uv)$ with coefficients $f_{lm}$. We introduce $\rrf{\uvk}$ for a unit vector $\uvk\in\Cm$ ($\uvk\cdot\uvk=1$) as
\be
\rrf{\uvk}f(\uv)=\sum_{l=0}^{\infty}\sum_{m=-l}^lf_{lm}Y_{lm}(\uv;\uvk),
\ee
where \cite{Markel04}
\be
Y_{lm}(\uv;\uvk)=\sum_{m'=-l}^le^{-im'\va_{\uvk}}d_{m'm}^l(\vth_{\uvk})Y_{lm'}(\uv).
\ee
Here, $\va_{\uvk},\vth_{\uvk}$ are the azimuthal and polar angles of $\uvk$ and $d_{m'm}^l$ are the Wigner $d$-matrices. We choose the branch cut of the square root function from $0$ to $\infty$, so $0\le\mathop{\mathrm{arg}}(\sqrt{z})<\pi$ for arbitrary $z\in\Cm$. That is, by $\rrf{\uvk}$, we measure angles in $f(\uv)$ in the reference frame rotated so that the $z$-axis lies in the direction of $\uvk$.

Indeed, $\nu$ be an eigenvalue or in the continuous spectrum. We take the specific form of $\uvk=\uvk(\nu,\vv{q})$ given below, which depends on $\nu$ and $\vv{q}\in\Rm^2$.
\be
\uvk(\nu,\vv{q})=\left(-i\nu\frac{\vv{q}}{\mu_t},\;\hat{k}_z(\nu q)\right),
\quad
\hat{k}_z(\nu q)=\sqrt{1+(\nu q/\mu_t)^2},
\ee
where $q=|\vv{q}|$. We note that $\uvk(\nu,\vv{q})\cdot\uvk(\nu,\vv{q})=1$. We obtain
\be
\va_{\uvk}=\left\{\begin{aligned}
\va_{\vv{q}}+\pi
&\quad\mbox{for}\;\nu>0,
\\
\va_{\vv{q}}
&\quad\mbox{for}\;\nu<0,
\end{aligned}\right.
\ee
where $\va_{\vv{q}}$ is the polar angle of $\vv{q}$, and
\be
\cos\vth_{\uvk}=\uvk\cdot\hvv{z}=\hat{k}_z,\quad
\sin\vth_{\uvk}=\sqrt{1-\cos^2\vth_{\uvk}}=i|\nu q|.
\ee
We note that $d_{m'm}^l(\vth_{\uvk})$ depends on $q$ but is independent of $\va_{\vv{q}}$. Hence we write
\be
d_{m'm}^l(\vth_{\uvk})=d_{m'm}^l[i\tau(\nu q)].
\ee

We define
\be
\begin{aligned}
\Psi_{\nu_j(M)}^M(\uv,\vv{q}_0)&=
\rrf{\uvk(\nu_j(M),\vv{q}_0)}\Phi_{\nu_j(M)}^M(\uv),
\\
\Psi_{\nu}^M(\uv,\vv{q}_0)&=
\rrf{\uvk(\nu,\vv{q}_0)}\Phi_{\nu}^M(\uv).
\end{aligned}
\ee

\subsection*{B. 2. Ballistic subtraction}

As was done in Appendix \ref{rte}, we consider $I_s$ by subtracting $I_b$ from $I$. Let us introduce the particular solution $I_p$ as
\be
\begin{aligned}
\left(\uv\cdot\nabla+\mu_a+\mu_s\right)I_p(\vv{r},\uv)
&=
\mu_s\int_{\Sm^2}p(\uv,\uv')I_p(\vv{r},\uv')\,d\uv'
\\
&+
\Theta(z)S(\vv{r},\uv),
\\
&\quad (\vv{r},\uv)\in\Rm^3\times\Sm^2,
\end{aligned}
\ee
where $\Theta(\cdot)$ is the step function. Then we can calculate $I_s$ as
\be
I_s=I_p+\psi,
\ee
where $\psi$ satisfies
\be
\left\{\begin{aligned}
\left(\uv\cdot\nabla+\mu_a+\mu_s\right)\psi(\vv{r},\uv)=
\mu_s\int_{\Sm^2}p(\uv,\uv')\psi(\vv{r},\uv')\,d\uv',
\\
(\vv{r},\uv)\in\Omega\times\Sm^2,
\\
\psi(\vv{r},\uv)=R_{\mathfrak{n}}(\uv\cdot\hvv{z})\psi(\vv{r},\uv_R)+I_{\rm inc}^{(p)}(\vv{r},\uv),
\\
(\vv{r},\uv)\in\Gamma_-
\end{aligned}\right.
\ee
with
\be
I_{\rm inc}^{(p)}(\vv{r},\uv)=
R_{\mathfrak{n}}(\uv\cdot\hvv{z})I_p(\vv{r},\uv_R)-I_p(\vv{r},\uv),
\quad\vv{r}\in\pp\Omega.
\ee

\subsection*{B. 3. Particular solution}

To find $I_p(\vv{r},\uv)$, we write
\be
I_p(\vv{r},\uv)=
\mu_se^{i\vv{q}_0\cdot\vs{\rho}}e^{-\mu_tz}\Theta(z)
\sum_{l=0}^{l_{\rm max}}\sum_{m=-l}^l\eta_{lm}e^{-im\va_{\vv{q}_0}}Y_{lm}(\uv).
\ee
By multiplying $Y_{lm}^*(\uv)$ on both sides of the RTE for $I_p$ and integrating over $\uv\in\Sm^2$, we arrive at the following linear system which determines $\eta_{lm}$.
\be
\begin{aligned}
&
\sum_{l'm'}\Biggl[
\frac{iq_0}{2}\Biggl(
-\delta_{m',m-1}\delta_{l',l-1}\sqrt{(l+m)(l+m-1)}
\\
&
+\delta_{m',m-1}\delta_{l',l+1}\sqrt{(l'-m')(l'-m'-1)}
\\
&+
\delta_{m',m+1}\delta_{l',l-1}\sqrt{(l-m)(l-m-1)}
\\
&
-\delta_{m',m+1}\delta_{l',l+1}\sqrt{(l'+m')(l'+m'-1)}
\Biggr)
\\
&-
\mu_t\delta_{m'm}\left(\delta_{l',l-1}\sqrt{l^2-m^2}
+\delta_{l',l+1}\sqrt{(l')^2-m^2}\right)
\\
&+
\delta_{m'm}\delta_{l'l}(2l+1)\sigma_l
\Biggr]\frac{\eta_{l'm'}}{\sqrt{2l'+1}}
\\
&=
\delta_{m0}\frac{(2l+1)g^l}{\sqrt{4\pi}}.
\end{aligned}
\ee

Suppose that the light in direction $\uv\in\Sm^2_-$is detected at $\vv{r}_d\in\pp\Omega$. Here, $\Sm^2_-$ denotes the set of unit vectors in outgoing directions. We have
\be
I_p(\vv{r}_d,\uv)=
\mu_se^{i\vv{q}_0\cdot\vs{\rho}}\sum_{l=0}^{l_{\rm max}}\sum_{m=-l}^l\eta_{lm}e^{-im\va_{\vv{q}_0}}Y_{lm}(\uv).
\ee

\subsection*{B. 4. General solution}
\label{generalsol}

Since the scattering phase function $p(\uv,\uv')$ only depends on $\uv\cdot\uv'$, we can rewrite (\ref{phasefunc}) as
\be
p(\uv,\uv')=
\sum_{l=0}^{l_{\rm  max}}\sum_{m=-l}^l{\rm g}^lY_{lm}(\uv;\uvk)Y_{lm}^*(\uv';\uvk),
\ee
for arbitrary $\uvk=\uvk(\nu,\vv{q})$. We note that
\be
\begin{aligned}
Y_{lM}^*(\uv;\uvk)
&=
\rrf{\uvk}Y_{lM}^*(\uv)
\\
&=
\sum_{m=-l}^le^{im\va_{\uvk}}d_{mM}^l(\vth_{\uvk})Y_{lm}^*(\uv).
\end{aligned}
\ee
Let us express the eigenmodes as
\be
\begin{aligned}
&
\psi_{\nu}(\vv{r},\uv,\vv{q})
\\
&=
\sum_{l=0}^{l_{\rm max}}\frac{1}{\sqrt{\sigma_l}}\braket{l}{\phi_n(M)}Y_{lM}(\uv;\uvk)e^{-\mu_t\uvk\cdot\vv{r}/\nu}
\\
&=
e^{i\vv{q}\cdot\vs{\rho}}e^{-\mu_t\hat{k}_z(\nu q)z/\nu}
\\
&\times
\sum_{l=0}^{l_{\rm max}}\frac{\braket{l}{\phi_{\nu}}}{\sqrt{\sigma_l}}
\sum_{m=-l}^le^{-im\va_{\uvk}}d_{mM}^l(\vth_{\uvk})Y_{lm}(\uv).
\end{aligned}
\ee
We substitute the above $\psi_{\nu}(\vv{r},\uv,\vv{q})$ in the homogeneous equation of the RTE. By using $l'=0,\dots,l_{\rm max}$ and $m'=0,\pm1,\dots,\pm l'$, we obtain
\be
\begin{aligned}
&
\sum_{l'=0}^{l_{\rm max}}\sum_{m'=-l'}^{l'}\frac{\braket{l'}{\phi_n(m')}}{\sqrt{\sigma_{l'}}}Y_{l'm'}(\uv;\uvk)\left(-\frac{\uv\cdot\uvk}{\nu}+1\right)\mu_t
\\
&=
\mu_s\sum_{l'=0}^{l_{\rm max}}\sum_{m'=-l'}^{l'}{\rm g}^{l'}\frac{\braket{l'}{\phi_n(m')}}{\sqrt{\sigma_{l'}}}Y_{l'm'}(\uv;\uvk).
\end{aligned}
\ee
By rotating the reference frame in the inverse direction, we arrive at
\be
\begin{aligned}
&
\sum_{m'=-l_{\rm max}}^{l_{\rm max}}\sum_{l'=|m'|}^{l_{\rm max}}\frac{\braket{l'}{\phi_n(m')}}{\sqrt{\sigma_{l'}}}Y_{l'm'}(\uv)\left(-\frac{\cos\vth}{\nu}+1\right)\mu_t
\\
&=
\mu_s\sum_{m'=-l_{\rm max}}^{l_{\rm max}}\sum_{l'=|m'|}^{l_{\rm max}}{\rm g}^{l'}\frac{\braket{l'}{\phi_n(m')}}{\sqrt{\sigma_{l'}}}Y_{l'm'}(\uv).
\end{aligned}
\ee
By multiplying $Y_{lm}^*(\uv)$ ($-(l_{\rm max}-1)\le m\le l_{\rm max}-1$, $|m|\le l\le l_{\rm max}$) on both sides and integrating over $\uv\in\Sm^2$, we obtain
\be
\begin{aligned}
&
\sum_{l'=|m|}^{l_{\rm max}}\left(b_{l+1}(m)\delta_{l+1,l'}+b_l(m)\delta_{l-1,l'}\right)
\braket{l'}{\phi_n(m)}
\\
&=
\frac{\nu_n(m)}{\mu_t}\braket{l}{\phi_n(m)}.
\end{aligned}
\ee
In the above equation we wrote $\nu=\nu_n(m)$. We see that $\frac{\nu}{\mu_t}=\nu_n(M)/\mu_t$ and $\ket{\phi_n(M)}$ are eigenvalues and eigenvectors of the following matrix-vector equation \cite{Markel04,Panasyuk-etal06}.
\be
B(M)\ket{\phi_n(M)}=\frac{\nu_n(M)}{\mu_t}\ket{\phi_n(M)},
\ee
where $M=0,\pm1,\dots,\pm (l_{\rm max}-1)$ and matrix $B(M)\in\Rm^{(l_{\rm max}-|M|+1)\times(l_{\rm max}-|M|+1)}$ is a tridiagonal matrix whose elements are given by
\be
\{B(M)\}_{ll'}=b_l(M)\delta_{l',l-1}+b_{l'}(M)\delta_{l',l+1}
\label{Bmatrix}
\ee
for $|M|\le l,l'\le l_{\rm max}$. We used the notation such that $\bra{l}B(M)\ket{l+1}=\bra{l+1}B(M)\ket{l}=b_{l+1}(M)$. These $\nu_n(M)$ are approximate eigenvalues and discretized values of the continuous spectrum of Case's $\nu$ \cite{Garcia-Siewert89,Machida15}. We note that for each pair of $\nu_n(M)$ and $\braket{l}{\phi_n(M)}$ ($l=|M|,\dots,l_{\rm max}$), there exists a pair of eigenvalue $-\nu_n(M)$ and eigenvector $(-1)^l\braket{l}{\phi_n(M)}$ \cite{Markel04}. In order for the specific intensity $\psi(\vv{r},\uv)$ to vanish as $z\to\infty$, we take only $\lfloor(l_{\rm max}-|M|+1)/2\rfloor$ eigenvalues and eigenvectors such that
\be
\nu_n(M)>0,\quad
n=1,2,\dots,\left\lfloor\frac{l_{\rm max}-|M|+1}{2}\right\rfloor.
\ee

From the point of view of the singular eigenfunction, the method of rotated reference frames is the spherical-harmonic expansion of the singular eigenfunction \cite{Machida14,Machida15,Machida19}:
\be
\Phi_{\nu}^m(\uv)\approx\sum_{l=|m|}^{l_{\rm max}}\xi_l^m(\nu)Y_{lm}(\uv).
\ee
Using $\braket{\phi_n(M)}{\phi_n(M)}=1$ and $\int_{\Sm^2}\mu|\Phi_{\nu}^m(\uv)|^2\,d\uv=2\pi\mathcal{N}^m(\nu)$ with the normalization factor $\mathcal{N}^m(\nu)$ from one-dimensional transport theory, we find
\be
\xi_l^m(\nu)=
\sqrt{\frac{2\pi\mu_t\mathcal{N}^m(\nu)}{\nu\sigma_l}}\braket{l}{\phi_n(m)}.
\ee
Furthermore we note that $\braket{l}{\phi_{-\nu}(M)}=(-1)^l\braket{l}{\phi_{\nu}(M)}$ \cite{Markel04}.

The specific intensity $\psi(\vv{r},\uv)$ is given by the superposition of eigenmodes $\psi_{\nu}(\vv{r},\uv,\vv{q})$ with separation constant $\nu$ as
\be
\begin{aligned}
&
\psi(\vv{r},\uv)=
\frac{1}{(2\pi)^2}\sum_{\nu>0}\int_{\Rm^2}C_{\nu}(\vv{q})\psi_{\nu}(\vv{r},\uv,\vv{q})\,d\vv{q}
\\
&=
\frac{1}{(2\pi)^2}\sum_{\nu>0}\int_{\Rm^2}C_{\nu}(\vv{q})e^{i\vv{q}\cdot\vs{\rho}}\sum_{l=0}^{l_{\rm max}}\sum_{m=-l}^l\frac{\braket{l}{\phi_{\nu}}}{\sqrt{\sigma_l}}(-1)^m
\\
&\times
e^{-im\va_{\vv{q}}}d_{mM}^l[i\tau(\nu q)]Y_{lm}(\uv)
e^{-\mu_t\hat{k}_z(\nu q)z/\nu}\,d\vv{q},
\end{aligned}
\ee
where $C_{\nu}(\vv{q})$ is determined later from the boundary condition. We note that
\be
\begin{aligned}
I_{\rm inc}^{(p)}(\vv{r},\uv)
&=
\mu_se^{i\vv{q}_0\cdot\vs{\rho}}\sum_{l=0}^{l_{\rm max}}\sum_{m=-l}^l\eta_{lm}e^{-im\va_{\vv{q}_0}}
\\
&\times
\left(R_{\mathfrak{n}}(\uv\cdot\hvv{z})(-1)^{l+m}-1\right)Y_{lm}(\uv).
\end{aligned}
\ee

Let us find $C_{\nu}(\vv{q})$. As was done in \cite{Machida-etal10}, we introduce
\be
\begin{aligned}
\mathcal{B}_{ll'}^m(\mathfrak{n})
&=
\int_{\Sm^2_+}R_{\mathfrak{n}}(\cos\vth)Y_{l'm}(\uv)Y_{lm}^*(\uv)\,d\uv
\\
&=
\frac{1}{2}\sqrt{\frac{(2l+1)(2l'+1)(l-m)!(l'-m)!}{(l+m)!(l'+m)!}}
\\
&\times
\int_0^1R_{\mathfrak{n}}(\mu)P_l^m(\mu)P_{l'}^m(\mu)\,d\mu.
\end{aligned}
\ee
Note that $\mathcal{B}_{ll'}^{-m}(\mathfrak{n})=\mathcal{B}_{ll'}^m(\mathfrak{n})$. Furthermore we let $\mathcal{B}_{ll'}^m(\infty)$ denote $\mathcal{B}_{ll'}^m(\mathfrak{n})$ with $R_{\mathfrak{n}}=1$. Let us take the Fourier transform for $\vs{\rho}$ and operate $\int_{\Sm^2_+}d\uv\,Y_{lm}^*(\uv)$ on the boundary condition. By introducing $f_{Mn}(q)$ as
\be
C_{\nu}(\vv{q})=(2\pi)^2f_{Mn}(q)\delta(\vv{q}-\vv{q}_0),
\ee
we obtain
\be
\begin{aligned}
&
\sum_{M=0}^{l_{\rm max}-1}
\sum_{n=1}^{\lfloor(l_{\rm max}-|M|+1)/2\rfloor}
\Biggl[\sum_{l'=\max(|m|,|M|)}^{l_{\rm max}}
\\
&
\left(\mathcal{B}_{ll'}^m(\infty)-(-1)^{l'+m}\mathcal{B}_{ll'}^m(\mathfrak{n})\right)\frac{\braket{l'}{\phi_n(M)}}{\sqrt{\sigma_{l'}}}
\\
&\times
\left(d_{mM}^{l'}(\vth_{\uvk})+(1-\delta_{M0})(-1)^Md_{m,-M}^{l'}(\vth_{\uvk})\right)\Biggr]
f_{Mn}(q)
\\
&=
\mu_s\sum_{l'=m}^{l_{\rm max}}\eta_{l'm}
\left((-1)^{l'}\mathcal{B}_{ll'}^m(\mathfrak{n})-(-1)^m\mathcal{B}_{ll'}^m(\infty)\right)
\end{aligned}
\ee
for $0\le l\le l_{\rm max}$, $0\le m\le l$. Note that equations for $m$ and $-m$ are the same, and hence $f_{-M,n}(q)=(-1)^Mf_{Mn}(q)$. Due to the fact that associated Legendre polynomials satisfy three-term recurrence relations, linearly independent equations are extracted from the above equations if equations with $l=m+1+2\alpha$ ($\alpha=0,1,\dots,\lfloor(l_{\rm max}-m-1)/2\rfloor$) are taken for $m=0,1,\dots,l_{\rm max}-1$.

Finally, we obtain
\be
\psi(\vv{r},\uv)=
e^{i\vv{q}_0\cdot\vs{\rho}}\sum_{l=0}^{l_{\rm max}}\sum_{m=-l}^l
(-1)^me^{-im\va_{\vv{q}_0}}Y_{lm}(\uv)K_{lm}(q_0,z),
\label{psisol}
\ee
where
\be
\begin{aligned}
&
K_{lm}(q_0,z)=
\sum_{M=-(l_{\rm max}-1)}^{l_{\rm max}-1}
\sum_{n=1}^{\lfloor(l_{\rm max}-|M|+1)/2\rfloor}
f_{Mn}(q_0)
\\
&\times
\frac{\braket{l}{\phi_n(M)}}{\sqrt{\sigma_l}}
d_{mM}^l[i\tau(\nu_n(M)q_0)]e^{-\mu_t\hat{k}_zz/\nu_n(M)}.
\end{aligned}
\ee

Thus,
\be
\begin{aligned}
J_+(\vv{r}_d)
&=
\int_{\Sm^2_-}(\cos\vth)I_p(\vv{r}_d,\uv)\,d\uv+\int_{\Sm^2_-}(\cos\vth)\psi(\vv{r}_d,\uv)\,d\uv
\\
&=
-e^{i\vv{q}_0\cdot\vs{\rho}_d}A_{\rm RTE}(q_0),
\end{aligned}
\label{mrrfJ2}
\ee
where
\be
\begin{aligned}
A_{\rm RTE}(q_0)
&=
\sqrt{\pi}\sum_{l=0}^{l_{\rm max}}(-1)^l
\sqrt{2l+1}\left(\int_0^1\mu P_l(\mu)\,d\mu\right)
\\
&\times
\left(\mu_s\eta_{l0}+K_l(q_0)\right).
\end{aligned}
\label{Atheor2}
\ee
Here,
\be
\begin{aligned}
K_l(q_0)
&=
K_{l0}(q_0,0)=
\sum_{M\ge0,n}f_{Mn}(q_0)\frac{\braket{l}{\phi_n(M)}}{\sqrt{\sigma_l}}
\\
&\times
\Bigl(d_{0M}^l[i\tau(\nu_n(M)q_0)]
\\
&+
(1-\delta_{M0})(-1)^Md_{0,-M}^l[i\tau(\nu_n(M)q_0)]\Bigr).
\\
\end{aligned}
\label{mrrfK}
\ee
Note that $\int_0^1\mu P_1(\mu)\,d\mu=\frac{1}{3}$, $\int_0^1\mu P_l(\mu)\,d\mu=0$ if $l>1$ is odd, and when $l$ is even,
\be
\int_0^1\mu P_l(\mu)\,d\mu=
\frac{(-1)^{\frac{l}{2}+1}l!}{2^l(l-1)(l+2)\left[\left(\frac{l}{2}\right)!\right]^2}.
\ee


\end{document}